\documentclass[conference]{IEEEtran}
\IEEEoverridecommandlockouts
\usepackage{cite}
\usepackage{amsmath,amssymb,amsfonts}
\usepackage{algorithmic}
\usepackage{graphicx}
\usepackage{textcomp}
\usepackage{xcolor}
\usepackage{booktabs}
\def\BibTeX{{\rm B\kern-.05em{\sc i\kern-.025em b}\kern-.08em
    T\kern-.1667em\lower.7ex\hbox{E}\kern-.125emX}}
\begin{document}

\title{Programmable Multi-input Buck-Boost Converter for Photovoltaics Arrays\\
{\footnotesize}
\thanks{This work was supported by the CLEAN-Power (Compatibility and Low electromagnetic Emission Advancements for Next generation Power electronic systems) project at the Department of Energy, Aalborg University, Aalborg, Denmark, funded by Independent Research Fund Denmark (DFF).}
}

\author{\IEEEauthorblockN{Zhongting Tang, Yi Zhang, and Pooya Davari}
\IEEEauthorblockA{\textit{AAU Energy, Aalborg University, Aalborg 9220, Denmark} \\
Email: zta@energy.aau.dk, yiz@energy.aau.dk, pda@energy.aau.dk}

}

\maketitle

\begin{abstract}
This paper proposes a programmable multi-input buck-boost structure method, which can enhance the operation tolerance for the PV array under extremely harsh climatic conditions. The proposed structure based on a traditional two switches buck-boost converter can connect PV panels in parallel and cascade flexibly, and also enable the individual operation of each PV panel. The active switches can be programmed to change the connection structures as well as achieve the maximum power point track of PV panels simultaneously. The paper presents the programming method for an exemplified scalable structure converter for two PV panels. The simulation has been established in MATLAB/Simulink to validate the performance of the proposed converter in terms of multiplexing function, wide operating range of PV panels, and low switching stress. 
\end{abstract}


\section{Introduction}\label{sec:Intro}
Solar energy continued to lead capacity expansion in renewables, e.g., wind, hydropower, wave, and so on~\cite{b1}. In 2022, the renewable generation capacity increased by 295 GW, where solar energy occupied the main part with a massive increase of 192 GW. It is estimated that the energy consumption of the world will reach 27~TW/yr per annum in 2050, and solar resources dwarfed all other renewables by orders of magnitude~\cite{b2}. Regarding carbon neutrality promotion of solar energy, the waste of PV power systems should be considered in the carbon emission. Therefore, further improving the utilized efficiency is a solution to increase the benefit of carbon neutrality from solar energy~\cite{b3}.

\begin{figure}[!t]\centering
	\includegraphics[width=1\columnwidth]{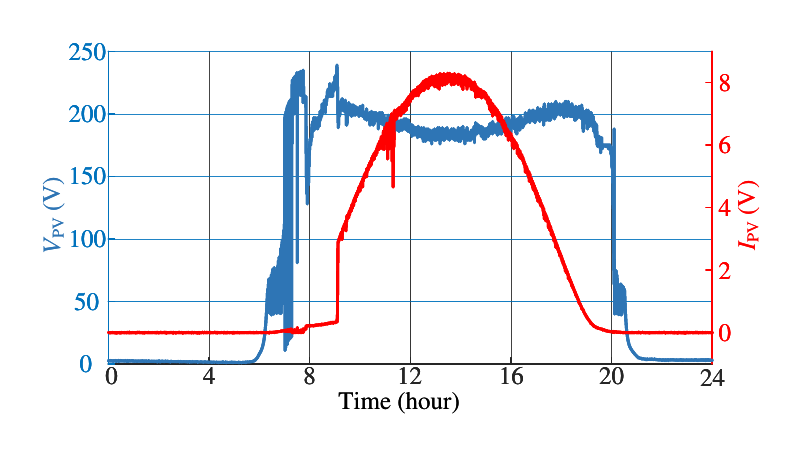}
	\vspace{-0.3cm}
	\caption{Voltage and current behaviors of the PV panels (i.e., Solvis SV36-150, 12 panels in series) under a daily mission profile, where $V_\text{PV}$ and $I_\text{PV}$ are the voltage and current of one PV array under the MPPT operation, respectively.}
	\vspace{-0.3cm}
	\label{fig1}
\end{figure}

In addition to the uncertainty of power generation, the maximum power point track (MPPT) voltage and current of PV panels have inherent characteristics~\cite{b4}, as shown in Fig.~\ref{fig1}. The mission profile of the voltage and current behavior of the PV panel in Fig.~\ref{fig1} can indicate that 1) the voltage is relatively high and the current is relatively low in the morning and twilight(i.e., due to the low irradiation and the low temperature), and 2) the voltage is low and the current is high at noon (i.e., because of the high irradiation and the high temperature). The wide range operation of the MPPT voltage and current thus should be implemented in first stage converters. Conventional buck-boost converters are widely employed thanks to their simplicity, but the disadvantages are high switching power losses~\cite{b5}. To alleviate this problem, many techniques have been developed in prior-art research, such as isolated, non-isolated, cascaded, switched, flying capacitor, coupled inductor-based DC-DC converters, and soft switching techniques~\cite{b5,b6,b7}. However, all these converters have some limitations in the context of high-gain applications.

Besides, the mismatched atmospheric conditions of the PV panels always cause the barrel effect, i.e., the modules operate at the least maximum power point (MPP) among all PV panels~\cite{b8}. The multiport DC-DC converter is a solution to solve this problem~\cite{b8,b9}. For instance, PV panels can maintain in cascade for the same atmospheric conditions, and operate individually during mismatched conditions ~\cite{b8}. Among different multiport DC-DC converters, the benchmark performance relates to switching stresses (both the voltage and the current), positive output voltage, high efficiency, high power density, flexible control, and so on. For instance, a family of multiport converters has been derived based on the two-switch buck-boost converter, which is a simplified cascade connection of the buck and boost converters~\cite{b9}. However, those multiport converters only connect the input sources in parallel, which still have to outage or lose much energy when the weather is extremely harsh (e.g., the MPPT voltage is too low and the current is too high). To further improve the multiplexing and the benchmark performance, advanced multiport converters should be developed with flexible cascade and parallel connections of PV panels.

This paper proposes a novel approach to program multi-input buck-boost converters with any number of ports scaling in series and parallel. Therefore, PV panels can be permuted and combined flexibly according to different atmosphere conditions, enhancing the wide MPPT voltage and current operation range without compromising the switching stresses. The rest is organized as follows. The main scalable idea of programming multi-input buck-boost converters is proposed as well as a two inputs buck-boost converter topology in Section II. In Section III, the typical two inputs topology of the proposed family converters is analyzed in terms of connection structures, MPPT operation scheme and the corresponding control strategy. Simulation results are presented to validate the different MPPT operations in Section IV. Section V gives the final conclusion.  
   
\begin{figure}[!t]\centering
	\includegraphics[width=0.9\columnwidth]{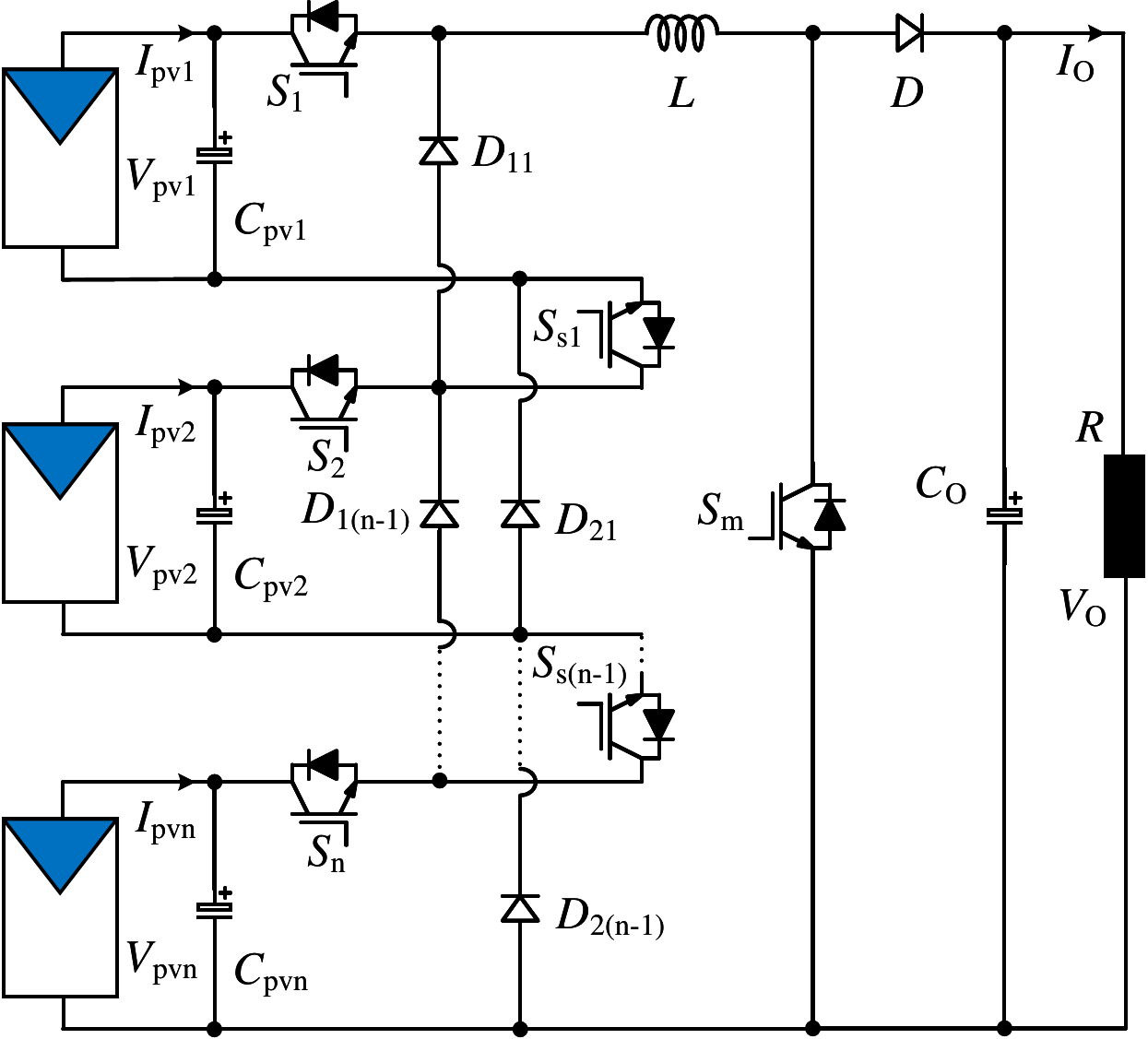}
	\caption{Proposed programable concept for multi-input buck-boost converters for PV panels.}
	\vspace{-0.3cm}
	\label{fig2}
\end{figure}

\section{Proposed Programable Multi-input Converter Structure}

The schematic of the proposed programmable multi-input converter concept is depicted in Fig.~\ref{fig2}. The input side is composed of multiplexing-commutated switches ($S_\text{1$\sim$n}$),  corresponding input capacitors $C_\text{pv1$\sim$n}$, and module bypass diodes ($D_\text{1(1$\sim$n-1)}$ and $D_\text{2(1$\sim$n-1)}$). Active switches $S_\text{s1$\sim$s(n-1)}$ control the connection structure to be cascade or parallel, and the input switches $S_\text{1$\sim$n}$ control the individual operations of each channel. Besides, the boost stage consists of a main switch $S_\text{m}$, a filter inductor $L$, a clamping diode $D$, and an output capacitor $C_\text{O}$. A load of the resistor $R$ is set here to validate the output power in the connection structures under different atmospheric conditions of PV panels. $V_\text{pv1$\sim$n}$ and $I_\text{pv1$\sim$n}$ are the input voltage and current of each PV panel, and $V_\text{O}$ and $I_\text{O}$ are the output voltage and current, separately.

\begin{figure}[!t]\centering
	\includegraphics[width=0.9\columnwidth]{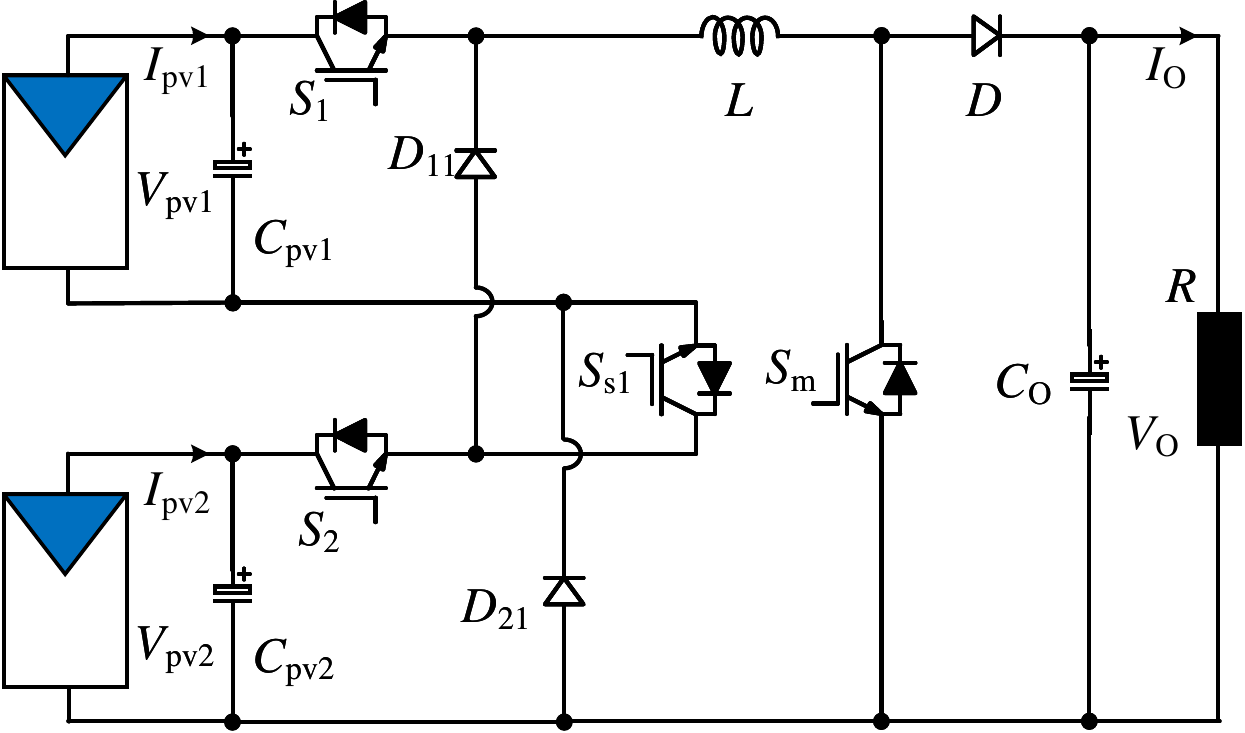}
	\caption{Exemplified two input ports buck-boost converter.}
	\vspace{-0.3cm}
	\label{fig3}
\end{figure}

This paper exemplified the proposed structure concept in a two inputs converter, i.e., $n=2$. That means four active switches and three diodes are included, as shown in Fig.~\ref{fig3}. The operation modes will be detailed as well as switching principles under different connections in the following.

\section{Analysis of Exemplified Converter}

The proposed multi-input converter has four different connection strcutures, as shown in Fig.\ref{fig4}. It includes the parallel connection (\textbf{Parallel}), the cascade connection (\textbf{Cascade}), the individual PV1 (\textbf{PV1}), and the individual PV2 (\textbf{PV2}), in which the operated switches are shown in Table.~\ref{tab_I}. 

\begin{table}[!t]
	\caption{Switches states in different structures.}
	\label{tab_I}
	\centering
	\begin{tabular}{lllllll}
		\midrule \midrule
		\textbf{Type} & $S_\text{1}$ & $S_\text{2}$ & $S_\text{s1}$ & $S_\text{m}$ & $D_\text{11}$ & $D_\text{21}$   \\
		\midrule
		\textbf{Parallel}   &  $+$  &  $+$  & $-$ & $+$ & $+$ & $+$\\
		\textbf{Cascade}    &  $+$  &  $+$  & $+$ & $+$ & $-$ & $-$\\
		\textbf{PV1}        &  $+$  &  $-$  & $-$ & $+$ & $+$ & $-$\\
		\textbf{PV2}        &  $-$  &  $+$  & $-$ & $+$ & $-$ & $+$\\		
		\midrule \midrule
	\end{tabular}
\\
\footnotesize             
\textbf{Note}: + means the switch/diode operates,\\
and $-$ means the switch/diode is OFF/blocked. \\
\vspace{-0.3cm}
\end{table}

\subsection{Connection structures} 
\begin{figure*}[!t]\centering
	\includegraphics[width=160mm]{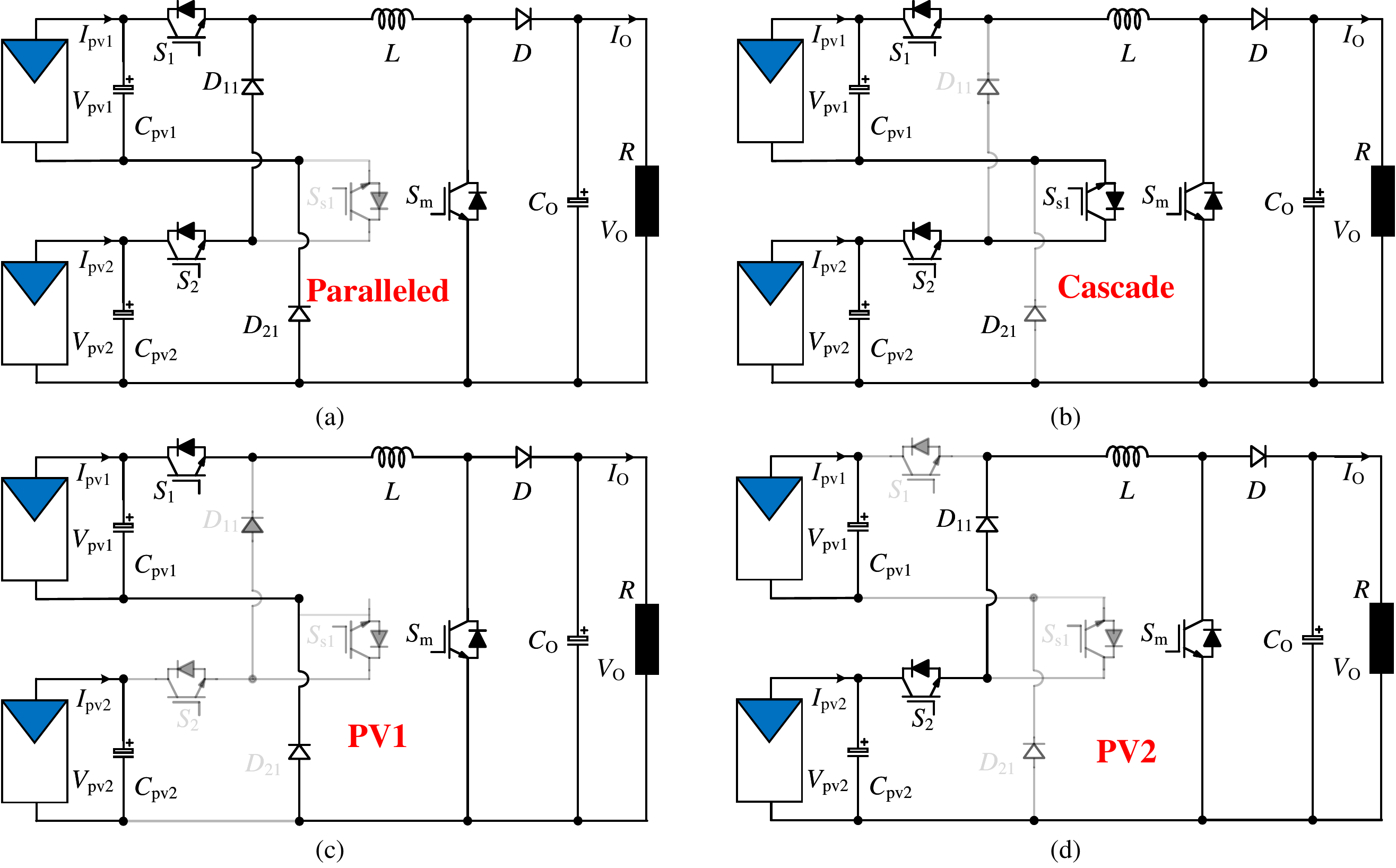}
	\caption{Connection structures of the proposed multi-input converter exemplified in Fig.~\ref{fig3}, (a) \textbf{Parallel} connection, (b) \textbf{Cascade} connection, (c) individual \textbf{PV1} connection, and (d) individual \textbf{PV2} connection.}
	\vspace{-0.3cm}
	\label{fig4}
\end{figure*}
1) \textbf{Parallel}: as shown in Fig.~\ref{fig4}(a),  $S_\text{s1}$ is OFF, $S_\text{1}$ and $S_\text{2}$ operate at a high switching frequency to transmit the energy of both PV1 and PV2 and perform the first DC-DC conversion. This parallel structure can be used in the atmosphere of high voltage and low current. As to the MPPT operation, this structure can turn on alternately within one switching cycle to enable the MPPT for each PV panel. Therefore, the duty cycles of $S_\text{1}$ and $S_\text{2}$ have limitations, where the total of duty cycles cannot exceed 1. $S_\text{1}$ and $S_\text{2}$ can also be ON simultaneously without considering the limitation. But it has the barrel effect of the MPPT operation mentioned previously.   

2) \textbf{Cascade}: when $S_\text{s1}$ is ON, and $S_\text{1}$ and $S_\text{2}$ switch at a high frequency, the proposed converter has a cascade connection (i.e., fig.~\ref{fig4}(b)). It is suitable for high current and low voltage conditions. The cascade connection can reduce the current stress for $L$ and $S_\text{m}$ in the converter. It should be mentioned that the cascade structure still has an inherent barrel effect for the MPPT due to the same current in each PV panel.

3) \textbf{PV1}: as depicted in Fig.~\ref{fig4}(c), $S_\text{1}$ is switching at a high frequency, and $S_\text{2}$ and $S_\text{s1}$ are both OFF.  The proposed converter transmits the energy from the PV1 individually, which often happens with the serious shielding of other PV panels (i.e., PV2 in the proposed two-input converter). 

4)  \textbf{PV2}: similarly, the PV2 operates individually in this connection, where $S_\text{2}$ is switching at a high frequency, and $S_\text{1}$ and $S_\text{s1}$ are OFF.

In the above connections, it should be noted that $S_\text{m}$ just executes the DC-DC conversion of the subsequent stage, which has no impact on the front structure options of the proposed multiple input converters.

\subsection{MPPT Operation Scheme}

\begin{figure}[!t]\centering
	\includegraphics[width=0.9\columnwidth]{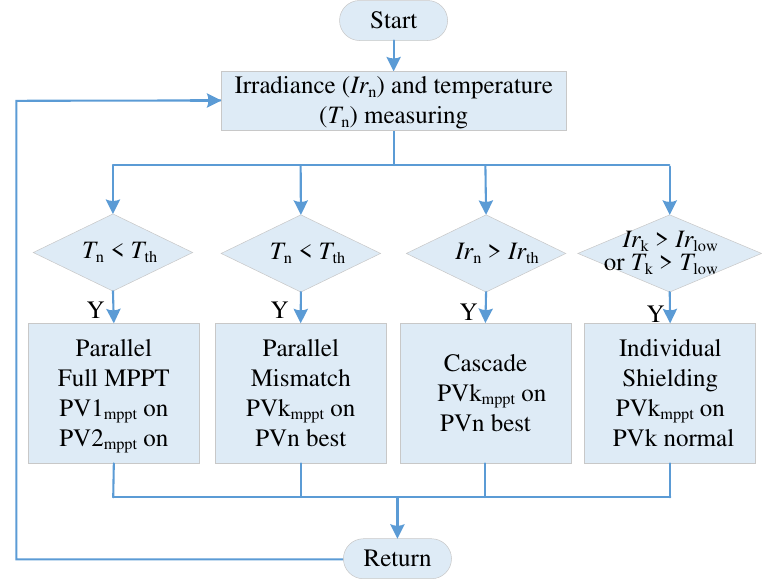}
	\caption{MPPT operation scheme under different connection structures.}
	\label{fig5}
\end{figure}

The MPPT operation schemes under different connection structures have been shown in Fig.~\ref{fig5}. The atmosphere conditions are used for changing the connection structures, where $Ir_\text{n}$ and $T_\text{n}$ represent the irradiation and the temperature conditions of each PV panel. The threshold values of the irradiation and the temperature (i.e., $Ir_\text{th}$ and $T_\text{th}$) set in the proposed scheme can be adjusted based on the specific atmosphere in different areas, which are used to change the connection structures of \textbf{Cascade} and \textbf{Parallel}. Besides, $Ir_\text{low}$ and $T_\text{low}$ are used for the judgment of the serious shielding for each PV panel. The MPPT operation schemes can be detailed in the following.

1) \textbf{Parallel with full MPPT}: as shown in Fig.~\ref{fig5}, when the atmosphere conditions of each panel are almost the same and the temperature meets $T_\text{n}<T_\text{th}$, both PV panels can enable the MPPT operation in the parallel connection. However, the duty cycle of each input switches $S_\text{1}$ and $S_\text{2}$ must satisfy the constraint that the sum does not exceed 1. Therefore, the MPPT operations for each panel are limited in a level.

2) \textbf{Parallel with Mismatch}: in addition to meeting $T_\text{n}<T_\text{th}$, when the atmosphere conditions of PV panels are mismatched, the PV panel with the best condition enables the MPPT operation. The input switches of the rest PV panels switch the same as the MPPT PV panel. There is no limitation to the duty cycles of each panel.

3) \textbf{Cascade}: when $Ir_\text{n}>Ir_\text{th}$, the current is high, and then those two PV panels are connected in cascade. Thus the voltage and current stress for the incutor $L$ and $S_\text{m}$ can be adjusted. As mentioned before, the barrel effect is inherent in the cascade connection. Thus, one PV panel with the best condition enables the MPPT operation, and the input switch of the rest PV panel follows the same commution.

4) \textbf{Individual}: when $Ir_\text{k}>Ir_\text{low}$ or $T_\text{k}>T_\text{low}$, the PV panel of $k$ is defined without shielding. Accordingly, the PV panel of $k$ operates the MPPT operation individually, and the rest are in the outage state.

\begin{figure}[!t]\centering
	\includegraphics[width=0.9\columnwidth]{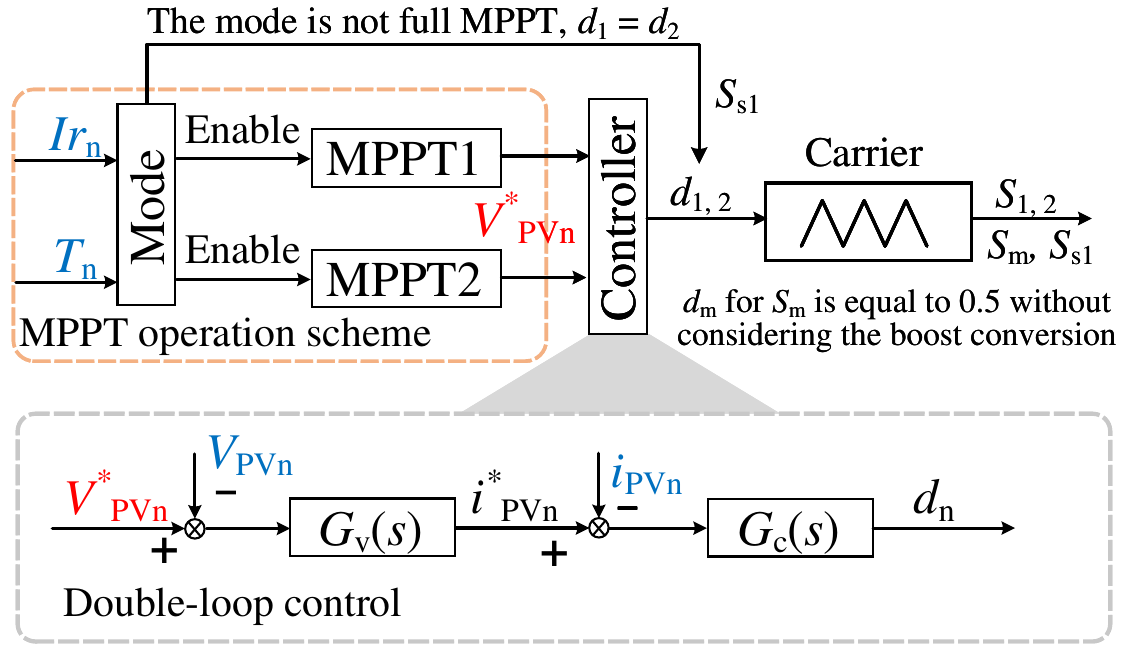}
	\caption{Control block of the two inputs converter, where $V_\text{PVn}$ and $i_\text{PVn}$ are the measured voltage and current of PVs, and $G_\text{v}(s)$ and $G_\text{c}(s)$ are the voltage loop controller and the current loop controller, respectively.}
	\label{fig6}
\end{figure}

Besides, the control block is presented in Fig.~\ref{fig6}. The measured $Ir_\text{n}$ and $T_\text{n}$ are used for the MPPT operation switching according to the scheme in Fig.~\ref{fig5}. Then, the Perturb and Observe (P\&O) algorithms are employed in the proposed scheme to achieve the reference voltage $V^*_\text{PVn}$ for PV panel $n$. The following is a double-closed-loop controller to generate the duty cycle $d_\text{n}$ for the corresponding input switch $S_\text{n}$. The outer voltage loop is a proportional-integral (PI) controller to generate the reference PV current $i^*_\text{PVn}$, then $d_\text{n}$ is generated through the current controller (i.e., PI controller). $S_\text{s1}$ is used for switching connection structures. Since only focusing on the flexible input structure, $d_\text{m}$ for $S_\text{m}$ is set to be 0.5, and connecting a resistor load.  

\section{Validation in Matlab/Simulink}
To validate the flexible connection structure of the proposed multiple-input converter, a prototype of the exemplified converter in Fig.~\ref{fig3} is built in Matlab/Simulink. The different connection structures are validated with two constant DC sources as the inputs, where the $V_\text{PV1}=V_\text{PV1}=$ 100 $V$. Then, each input is a string of 12 PV panels with parameters of the commercial PV module Solvis SV36-150~\cite{b10} to verify the proposed MPPT scheme. The load resistor is 20 $\Omega$. 

\begin{figure}[!t]\centering
	\includegraphics[width=0.9\columnwidth]{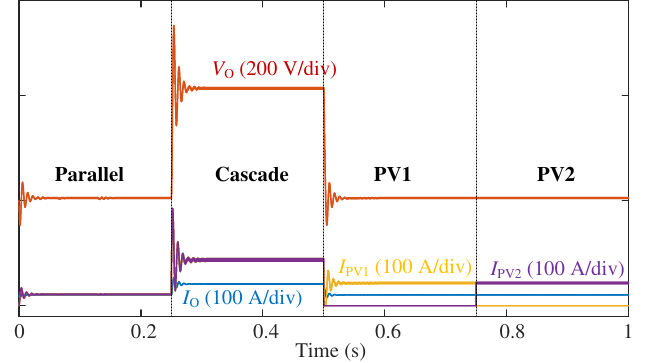}
	\caption{Performance of different connection structures under constant DC sources.}
	\label{fig7}
\end{figure}

The simulation result under different connection structures is shown in Fig.~\ref{fig7}, where $I_\text{PV1}$ and $I_\text{PV1}$ are the two input PV currents,  $I_\text{O}$ and $V_\text{O}$ are the load current and voltage, respectively. During $0 s$ to $0.25 s$, the two-input converter is in parallel connection, where the load voltage $V_\text{O}$ is 200 V, $I_\text{PV1}$, $I_\text{PV2}$ and $I_\text{O}$ are 10 A, respectively. The power is 2 KW at this casa. In the period from $0.25 s$ to $0.5 s$, the converter is in cascade connection, where $V_\text{O}$ is 400 $V$, $I_\text{PV1}$ and $I_\text{PV2}$ are 160 A, and $I_\text{O}$ is 40 A, respectively (i.e., the power is 32 KW). In the phase of $(0.5 s- 0.75 s)$ and $(0.75 s-1 s)$, the converter operates at individual connections PV1 and PV 2, seperately. $V_\text{O}$ is 200 V, $I_\text{O}$ is 10 A, and $I_\text{PV1}$/$I_\text{PV2}$ is 20 A/0 A (i.e., the power is 2 KW). The results can verify that the proposed multiple inputs converter has the multiplexing functions, i.e., parallel, cascade, and individual.   

\begin{figure*}[!t]\centering
	\includegraphics[width=165 mm]{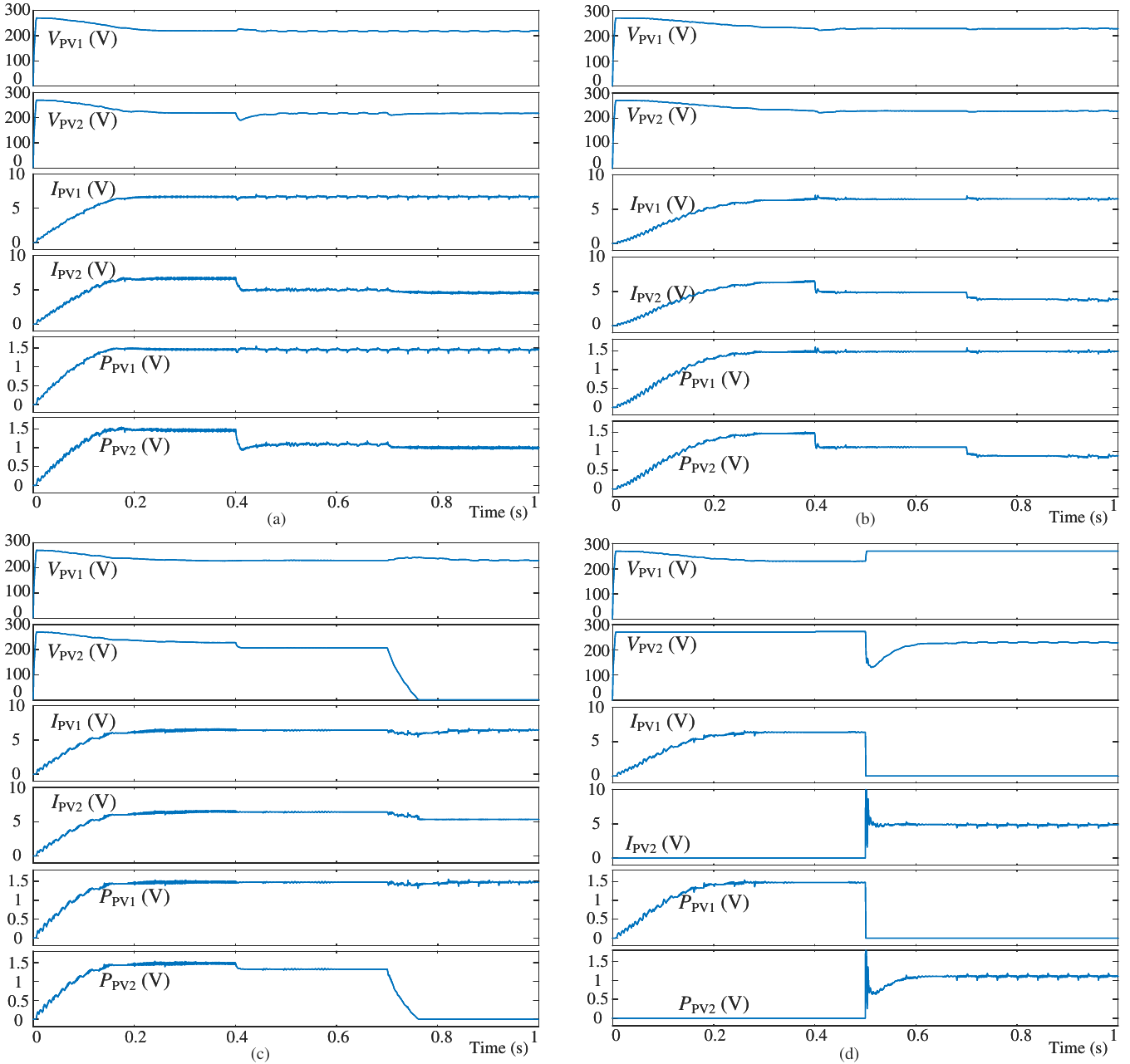}
	\caption{Simulation results of the proposed MPPT scheme, (a) Parallel connection with full MPPT operations, (b) Parallel connection with one PV mismatch, (c) Cascade connection with one PV mismatch, and (d) Individual MPPT operation.}\vspace{-0.3cm}
	\label{fig8}
\end{figure*}

Fig.~\ref{fig8} shows the performance of MPPT operations under different connection structures, where $V_\text{PV1}$ and $V_\text{PV2}$ are the two input PV voltages, $I_\text{PV1}$ and $I_\text{PV2}$ are the two input PV currents, and $P_\text{PV1}$ and $P_\text{PV2}$ are the power of two input PV panels. Fig.~\ref{fig8} (a) shows the simulation in the parallel connection with the MPPT operation for each PV panel. It can be seen the PV1 and the PV2 can act in MPPT operation even when PV2 happens the step changes of the irradiation and temperature at 0.4 s and 0.7 s, respectively. By comparison, PV2 only follow the same duty cycle as PV1 in the parallel connection with mismatch conditions, where PV1 has the MPPT operation and PV2 decreases the power at 0.7 s continuously in Fig.~\ref{fig8} (b). In the cascade connection of Fig.~\ref{fig8}(c), the MPPT operation can be achieved by PV1. PV2 follows the same duty cycle of PV1, where $V_\text{PV2}$ decreases to 0 when the mismatch is serious. Fig.~\ref{fig8} (d) shows the MPPT operation of each PV panel in the individual operation. The rest PV panel with serious shielding is in an outage state, where the corresponding current is 0. The simulation results validate the effectiveness of the proposed MPPT operation scheme in Fig.~\ref{fig5}.

In all, the simulation results validate that the proposed scalable multi-input buck-boost converter can achieve multiplexing function, wide operating range of PV panels, and low switching stress. Therefore, it can enhance high conversion efficiency from solar energy to electric power more flexibly along with the changeable weather conditions.

\section{Conclusion}
This paper proposed a programmable multiple-input buck-boost structure concept based on the two switches buck-boost converter. The analysis of the different connection structures and operation principles has been presented in an exemplified two inputs converter. Besides, the corresponding MPPT operation scheme as well as the system control method have been developed and described under different connection structures. Finally, the simulation verifies the multiplexing function and flexibility of the proposed scalable multi-input converter.


\begin{thebibliography}{00}
\bibitem{b1} IRENA, ``Renewable energy capacity statistics 2023,'' http://www.irena.org/publications, 2023.	
\bibitem{b2} R. P. Marc Perez, ``Update 2022-a fundamental look at supply side energy reserves for the
planet,'' \textit{Solar Energy Advances}, vol. 2, p. 100014, 2022.
\bibitem{b3} IEA, ``Pv-powered electric vehicle charging stations preliminary requirements and feasibil-ity conditions,'' https://iea-pvps.org/key-topics/pv-powered-electric-vehicle-charging-stations/, 2021.
\bibitem{b4} Z. Tang, Y. Yang, and F. Blaabjerg, ``Power electronics: The enabling technology for renewable
energy integration,'' \textit{CSEE Journal of Power and Energy Systems}, vol. 8, no. 1, pp. 39--52, Jan. 2022.
\bibitem{b5} W. Li and X. He, ``Review of Nonisolated High-Step-Up DC/DC Converters in Photovoltaic Grid-Connected Applications,'' \textit{IEEE Trans. Ind. Electron.}, vol. 58, no. 4, pp. 1239--1250, Apr. 2011.
\bibitem{b6} A. K. Singh, A. K. Mishra, K. K. Gupta, and Y. P. Siwakoti, ``High voltage gain bidirectional	dc-dc converters for supercapacitor assisted electric vehicles: A review,'' \textit{CPSS Transactions on Power Electronics and Applications}, vol. 7, no. 4, pp. 386--398, 2022.
\bibitem{b7} C. Yao, X. Ruan, X. Wang, and C. K. Tse, ''Isolated Buck-Boost DC/DC Converters Suitable for Wide Input-Voltage Range,'' \textit{IEEE Trans. Power Electron.}, vol. 26, no. 9, pp. 2599--2613, Sep. 2011.
\bibitem{b8} A. Hema Chander, L. K. Sahu, and S. Ghosh, ''Stand-alone multiple input photovoltaic in-
verter for maximum power extraction and voltage regulation under mismatched atmospheric
conditions,'' \textit{IET Renew. Power Gen.}, vol. 14, no. 9, pp. 1584--1595, 2020.
\bibitem{b9} H. Wu, J. Zhang, and Y. Xing, ``A Family of Multiport Buck-Boost Converters Based on DC-Link-Inductors (DLIs),'' \textit{IEEE Trans. Power Electron.}, vol. 30, no. 2, pp. 735--746, Feb. 2015.
\bibitem{b10} Datasheet, ``Model sv36,'' SOLVIS Photovoltaic Module, 2019.
\end{thebibliography}
\end{document}